# Electric field thermopower modulation analyses of the operation mechanism of transparent amorphous SnO$_2$ thin-film transistor

Dou-dou Liang[1,2,a)], Yu-qiao Zhang[2], Hai Jun Cho[2], and Hiromichi Ohta[2,a)]

AFFILIATIONS

[1]The Beijing Municipal Key Laboratory of New Energy Materials and Technologies, School of Materials Science and Engineering, University of Science and Technology Beijing, Beijing 100083, China

[2]Research Institute for Electronic Science, Hokkaido University, N20W10, Kita, Sapporo 001−0020, Japan

[a)]**Author to whom correspondence should be addressed:** liangdoudou1993@foxmail.com, hiromichi.ohta@es.hokudai.ac.jp

ABSTRACT

Transparent amorphous oxide semiconductors (TAOSs) based transparent thin-film transistors (TTFTs) with high field effect mobility ($\mu_{FE}$) are essential for developing advanced flat panel displays. Among TAOSs, amorphous (a-) $SnO_2$ has several advantages against current a-$InGaZnO_4$ such as higher $\mu_{FE}$ and being indium free. Although a-$SnO_2$ TTFT has been demonstrated several times, the operation mechanism has not been clarified thus far due to the strong gas sensing characteristics of $SnO_2$. Here we clarify the operation mechanism of a-$SnO_2$ TTFT by electric field thermopower modulation analyses. We prepared a bottom-gate top-contact type TTFT using 4.2-nm-thick a-$SnO_2$ as the channel without any surface passivation. The effective thickness of the conducting channel was ~1.7 ±0.4 nm in air and in vacuum, but a large threshold gate voltage shift occurred in different atmospheres; this is attributed to carrier depletion near at the top surface (~2.5 nm) of the a-$SnO_2$ due to its interaction with the gas molecules and the resulting shift in the Fermi energy. The present results would provide a fundamental design concept to develop a-$SnO_2$ TTFT.

Transparent amorphous oxide semiconductors (TAOSs) based transparent thin-film transistors (TTFTs) with high field effect mobility ($\mu_{FE}$) are essential components for advanced flat panel displays such as transparent organic light-emitting diode (OLED) displays and rollable OLED displays. Currently, amorphous (a-) $InGaZnO_4$ is widely applied as the TAOS[1-4] of the TFT channel of commercially available OLED displays; the optical bandgap of a-$InGaZnO_4$ is ~3 eV[5], transparent in the visible light region, and the $\mu_{FE}$ of a-$InGaZnO_4$ (~10 cm$^2$ V$^{-1}$ s$^{-1}$ [1,2]) is two orders of magnitude higher than that of previously used a-Si. However, the use of a-$InGaZnO_4$ must be reduced because the consumption of rare element such as indium (Clarke number: $1 \times 10^{-5}$ %) is not desirable for maintaining sustainable usage of resources. Therefore, development of post-a-$InGaZnO_4$ TAOSs showing high $\mu_{FE}$, which are composed of abundant elements, is crucial.

Among several indium-free TAOSs,[6-8] a-$SnO_2$ is a promising candidate for overcoming the issues with a-$InGaZnO_4$. Sn is one of abundant metal elements with the Clarke number[9] of Sn is $4 \times 10^{-3}$ %, 400 times larger than that of In. Moreover, $SnO_2$ TFT shows extremely high $\mu_{FE}$ >100 cm$^2$ V$^{-1}$ s$^{-1}$,[10-12] which is one order of magnitude higher than that of a-$InGaZnO_4$. In 2007, Dattoli *et al.*[10] reported that Ta-doped $SnO_2$ nanowire TFT exhibits uniform characteristics with average $\mu_{FE}$ exceeding 100 cm$^2$ V$^{-1}$ s$^{-1}$ at room temperature. In 2009, Sun *et al.*[11] fabricated Sb-doped $SnO_2$ nanocrystal TFTs at room temperature with $\mu_{FE}$ of 158 cm$^2$ V$^{-1}$ s$^{-1}$. In 2016, Shin *et al.*[12] reported that an extremely thin (< 4.5 nm) undoped $SnO_2$ TFT exhibited $\mu_{FE}$ of 150 cm$^2$ V$^{-1}$ s$^{-1}$ at room temperature in air. In these reports, bottom-gate top-contact type TFTs without any passivation showed high $\mu_{FE}$. Thus, the top surface of $SnO_2$ channel was exposed to the atmosphere.

SnO$_2$ is a well-known gas sensing material.[13,14] It is known that a 3−4-nm-thick depletion layer is formed at the SnO$_2$ surface[15,16]. Oxygen molecules in the ambient atmosphere are adsorbed on the surface, seize free electrons near the surface, and form a depletion layer. In a reducing atmosphere, parts of the seized electrons are released back to the depletion layer. Thus, the transistor characteristics of the bottom-gate top-contact type SnO$_2$ TFTs would be strongly affected by the gases surrounding the exposed channel surface due to their gas sensing property. For this reason, the operation mechanism of the bottom-gate top-contact type SnO$_2$ TFTs including the conduction band bending and the effective channel thickness have not been clarified thus far.

Here we clarify the conduction band bending and the effective channel thickness of 4.2-nm-thick a-SnO$_2$ based bottom-gate top-contact type TTFTs without any surface passivation. We analyzed the conduction band bending of the SnO$_2$ top surface by measuring the TFT characteristics in air and vacuum, and a large threshold gate voltage shift was observed; the Fermi energy in the carrier depletion region at the top surface (~2.5 nm) of the a-SnO$_2$ channel sensitively shifted with the changes in the gas atmosphere. The effective channel thickness ($t_{eff}$) was analyzed by the electric field thermopower ($S$) modulation method [17-21], and the $t_{eff}$ was ~1.7 ±0.4 nm in air and in vacuum. The present results would provide a fundamental design concept for developing a-SnO$_2$ TTFT.

The bottom-gate top-contact TTFTs were fabricated on 100-nm-thick ITO coated alkali-free glass (thickness: 0.7 mm, Corning® EAGLE XG®) substrates by pulsed laser deposition (PLD, KrF excimer laser, 10 Hz) technique. Firstly, a 300-nm-thick polycrystalline Y$_2$O$_3$ gate dielectric film (the dielectric permittivity, $\varepsilon_r$ = 20[22]) was deposited at room temperature. The fluence of the KrF laser was ~2 J cm$^{-2}$ pulse$^{-1}$ and the oxygen pressure was kept at 0.4 Pa during the deposition. Then, a 4.2-nm-thick SnO$_2$ film was deposited on the Y$_2$O$_3$/ITO bilayer laminate through a stencil mask at 300 °C [See Supplementary Figs. S1(a) and S1(b)]. The fluence of the KrF laser was ~0.3 J cm$^{-2}$ pulse$^{-1}$ and the oxygen pressure was kept at 1 Pa during the deposition. The optical bandgap ($E_g$) of the a-SnO$_2$ film was ~4.3 eV [Supplementary Fig. S1(c)], which was larger than that of bulk SnO$_2$ ($E_g$ ~3.6 eV).[23] This is most likely due to the quantum size effect.[24] Finally, 100-nm-thick ITO films, which were used as the source and drain electrodes (400 µm × 400 µm), were deposited at room temperature. The fluence of the KrF laser was ~0.9 J cm$^{-2}$ pulse$^{-1}$ and the oxygen pressure was kept at 3 Pa during the deposition. After these PLD processes, the device was annealed at 400 °C for 30 min in air. The channel length $L$ and the channel width $W$ of the resultant TFT were 200 and 400 µm, respectively. The resultant multiple layer was fully transparent in the visible light region (Supplementary Fig. S2).

The TTFT characteristics such as transfer characteristics ($I_d$–$V_g$) and output characteristics ($I_d$–$V_d$) were measured using a semiconductor device analyzer (B1500A, Agilent Co.) at room temperature. As shown in Fig. 2(a), the resultant TTFTs showed clear transfer ($I_d$–$V_g$) characteristics with the on-to-off current ratios of ~10$^5$. All the TFTs show clear pinch-off in the output characteristics [Supplementary Fig. S3], indicating that the TFT operation obeys the standard field-effect theory. It should be noted that the threshold gate voltage ($V_{th}$), which was evaluated by plotting $I_d^{0.5}$–$V_g$ relationship, was −14 V in air but shifted dramatically to −23 V in vacuum. The $\mu_{FE}$, calculated from $\mu_{FE} = g_m \cdot [(W/L) \cdot C_i \cdot V_d]^{-1}$, where $g_m$ is the transconductance $\partial I_d / \partial V_g$ and $C_i$ is capacitance per unit area ($C_i$ ~58 nF cm$^{-2}$),[22] was ~20 cm$^2$ V$^{-1}$ s$^{-1}$ in air and ~30 cm$^2$ V$^{-1}$ s$^{-1}$ in vacuum. These confirm that the channel conductance and the $V_{th}$ indeed can be modulated with the gases surrounding exposed a-SnO$_2$ channel. The SnO$_2$ TTFT characteristics were summarized in TABLE I.

The electric field modulated thermopower ($S$) was measured during the transfer characteristics measurements in the two gas atmospheres (air, vacuum) to analyze the $t_{eff}$. Details of the electric field modulated $S$ measurement are described elsewhere.[17-20] Figure 2(b) shows the changes in the $-S$ as a function of the effective gate voltage ($V_g-V_{th}$). The $S$ values were always negative, consistent with the fact that the SnO$_2$ film is $n$-type semiconductor. The absolute values of $S$ decrease gradually with increasing $V_g-V_{th}$ in both atmospheres due to an increase in the sheet carrier concentration ($n_s$), which was deduced from $n_s = C_i \cdot (V_g - V_{th}) \cdot e^{-1}$. Although small variation in $S$ was observed due to the gate leakage current ($I_g$), the observed $S$ could be used to analyze the $t_{eff}$ since the variation is less than 10%, and the difference in the air and vacuum atmospheres is noticeably clear.

In order to extract the $t_{eff}$, we plotted the $-S$ as a function of $n_s$ [Fig. 3(a)]. An almost linear relationship with a slope of ~120 µV K$^{-1}$ decade$^{-1}$ was observed in the $-S$ vs log $n_s$ plot when $n_s$ exceeded 2.5 × 10$^{12}$ cm$^{-2}$ in air, and a slope of ~84 µV K$^{-1}$ decade$^{-1}$ was observed in the same plot when $n_s$ exceeded 2.9 × 10$^{12}$ cm$^{-2}$ in vacuum. We also plotted the three-dimensional carrier concentration ($n_{3D}$) dependence of $-S$ measured from separately prepared SnO$_2$ thin films [Fig. 3(b), Table S1]. From the $-S$ vs log $n_{3D}$ relationship, the carrier effective mass ($m^*$) of the SnO$_2$ films was extracted to be 0.47 $m_0$ using the following equations:

$$s = \frac{k_B}{e} \left( \frac{\left(r+\frac{5}{2}\right) F_{\left(r+\frac{3}{2}\right)}(\eta)}{\left(r+\frac{3}{2}\right) F_{\left(r+\frac{1}{2}\right)}(\eta)} - \eta \right) \quad (1)$$

$$n_H = \frac{1}{eR_H} = \frac{8\pi(2m_d^* k_B T)^{\frac{3}{2}}}{3h^3} \frac{\left(r+\frac{3}{2}\right)^2 F^2_{\left(r+\frac{1}{2}\right)}(\eta)}{\left(2r+\frac{3}{2}\right) F_{\left(2r+\frac{1}{2}\right)}(\eta)} \quad (2)$$

$$F_n(\eta) = \int_0^\infty \frac{x^n}{1+e^{x-\eta}} dx \quad (3)$$

Where $F_n(\eta)$ is the $n$th order Fermi integral, $\eta$ is the reduced fermi energy, $r$ is the scattering factor, $h$ is the Plank constant, $k_B$ is the Boltzmann constant. Because acoustic phonon scattering is commonly the main scattering mechanism for this material, the scattering factor $r$ could be considered as -1/2. The S only depends on the slope of the electronic density of states at the Fermi level, which depends on $n_{3D}$. Since the $n_s$ measured from the transistor characteristics represents the projected $n_{3D}$ within the $t_{eff}$, we can directly compare the $-S$ vs log $n_s$ relationship with the $-S$ vs log $n_{3D}$ relationship. The slope of $-S$ vs log $n_{3D}$ plot [Fig. 3(b)] is comparable to that of the $-S$ vs log $n_s$ plot [Fig. 3(a)], indicating that the $E-k$ relation at the bottom of the conduction band is parabolic, and the $t_{eff}$ can be extracted as $n_s/n_{3D}$.

Several $S$-values around the lower $n_s$ did not follow the straight line, probably due to that the non-parabolic shaped tail states just below the original conduction band bottom.[18,22,25] The $S$ and $n_s$ were modulated from (–150 μV K$^{-1}$, 2 × 10$^{12}$ cm$^{-2}$) to (–80 μV K$^{-1}$, 1 × 10$^{13}$ cm$^{-2}$) in air and from (–110 μV K$^{-1}$, 2.9 × 10$^{12}$ cm$^{-2}$) to (–60 μV K$^{-1}$, 1.3 × 10$^{13}$ cm$^{-2}$) in vacuum with increasing positive electric field in the a-SnO$_2$ channel. The difference between the measurements in air and in vacuum is attributed to the gas sensing property of SnO$_2$, where O$_2$ molecules are adsorbed in air and released in vacuum. The $t_{eff} \equiv n_s/n_{3D}$ of the conducting a-SnO$_2$ channel were always ~1.7 ±0.4 nm, insensitive to the gas atmosphere [Fig. 3(c)].

Here, we would like to discuss the operation mechanism of a-SnO$_2$ TTFT [Fig. 4]. Without any $V_g$ application [Fig. 4(a)], the conduction band minimum (CBM) at the surface is lifted due to the adsorption of oxygen when the TTFT is exposed to air (higher oxygen atmosphere, black line). When the TTFT is exposed to vacuum, the CBM at the surface is lowered (lower oxygen atmosphere, red line). Due to the oxygen adsorption, a 2.5-nm-thick depletion layer, which is similar to the Debye length of SnO$_2$ (~3 nm)[26], is formed near the surface region. As a result, a conducting channel (~1.7 ±0.4 nm) remains at the Y$_2$O$_3$/a-SnO$_2$ interface, and the $n_s$ is ~4 × 10$^{12}$ cm$^{-2}$ ($n_{3D}$ ~2 × 10$^{19}$ cm$^{-3}$). Under positive $V_g$ application [Fig. 4(b)], the carriers accumulate at the Y$_2$O$_3$/a-SnO$_2$ interface, and the $E_F$ locates above the CBM. As a result, the 1.7-nm-thick 2D electron gas (2DEG) is formed at the Y$_2$O$_3$/a-SnO$_2$ interface. The $n_s$ increases up to ~1 × 10$^{13}$ cm$^{-2}$ ($n_{3D}$ ~4 × 10$^{19}$ cm$^{-3}$). The surface region does not change. Under negative $V_g$ application [Fig. 4(c)], the conduction electrons at the Y$_2$O$_3$/a-SnO$_2$ interface are completely depleted, showing off states in the transistor characteristics.

In summary, we have clarified the operation mechanism including the conduction band bending and the effective channel thickness ($t_{eff}$) of 4.2-nm-thick a-SnO$_2$ based bottom-gate top-contact type TTFTs without any surface passivation. We analyzed the conduction band bending of SnO$_2$ top surface by measuring the TFT characteristics in air as well as in vacuum and found the large threshold gate voltage shift; the Fermi energy in the carrier depletion region at the top surface (~2.5 nm) of the a-SnO$_2$ sensitively shifted depending on the gas atmospheres. We also analyzed the $t_\text{eff}$ ($\equiv n_\text{s}/n_\text{3D}$) by the electric field thermopower modulation method. The $t_\text{eff}$ was ~1.7 ±0.4 nm in air and in vacuum.

From the thickness of the a-SnO$_2$ film (4.2 nm) and the $t_\text{eff}$ (1.7 ±0.4 nm), the carrier depletion depth at the top surface of the a-SnO$_2$ film is estimated to be 2.5 nm, which is similar with the depletion length reported in other studies.[15,16] When a-SnO$_2$ bottom-gate top-contact TTFTs are exposed to air, oxygen gas would be adsorbed on the surface of SnO$_2$ film as electron accepting species, which is accompanied by the formation of a depletion layer inside the film. The present results may provide a fundamental design concept for utilizing a-SnO$_2$ TFT in device applications.

SUPPLEMENTARY MATERIAL

See the supplementary material for additional Crystallographic analyses of the SnO$_2$ thin film. Optical transmission of the bottom-gate top-contact a-SnO$_2$ TTFT. Transistor characteristics of the bottom-gate top-contact a-SnO$_2$ TTFT. Optical absorption spectrum of the polycrystalline Y$_2$O$_3$ thin film deposited on SiO$_2$ glass substrate.

ACKNOWLEDGMENTS

This research was supported by Grants-in-Aid for Innovative Areas (19H05791) and Scientific Research A (17H01314) from the JSPS. D.L. greatly appreciates the support from China Scholarships Council (201806460051). Y.Z. acknowledges the support from Grant-in-Aid for JSPS Fellows (19F1904909) from the JSPS. H.J.C. acknowledges the support from Nippon Sheet Glass Foundation for Materials Science and Engineering. H.O. acknowledges the support from the Asahi Glass Foundation and the Mitsubishi Foundation. A part of this work was also supported by Dynamic Alliance for Open Innovation Bridging Human, Environment, and Materials, and by the Network Joint Research Center for Materials and Devices.

TABLE I. On-ff current ratio, threshold gate voltage ($V_{th}$), subthreshold swing factor (*S.S.*), and field effect mobility ($\mu_{FE}$) for SnO$_2$ TTFT.

|        | ON/OFF       | $V_{th}$ (V) | *S.S.* (V decade$^{-1}$) | $\mu_{FEmax}$ (cm$^2$ V$^{-1}$ s$^{-1}$) |
|--------|--------------|--------------|--------------------------|-------------------------------------------|
| air    | ~ 10$^5$     | -14          | 0.65±0.02                | 20                                        |
| vacuum | ~ 10$^5$     | -23          | 1.57±0.2                 | 30                                        |

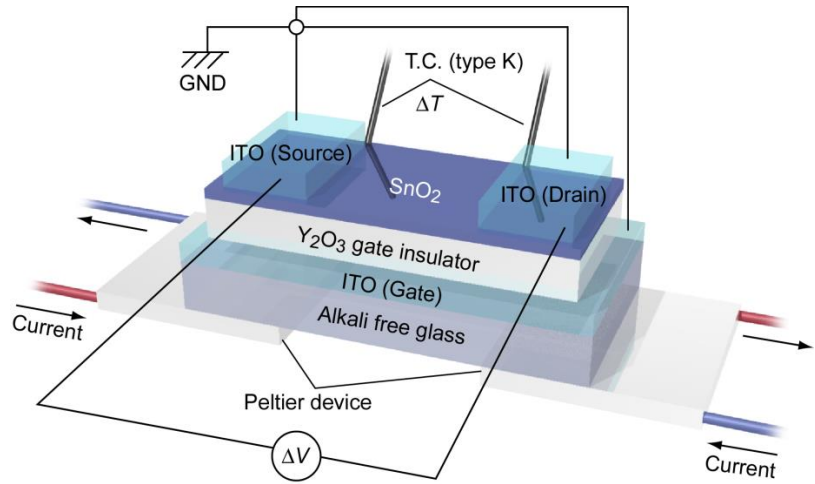

**FIG. 1** | Electric filed thermopower modulation measurement of the bottom-gate top-contact a-SnO$_2$ TTFT. The top surface of the 4.2-nm-thick SnO$_2$ channel is exposed to the air. The channel length $L$ is 200 μm and the channel width $W$ is 400 μm. The gate insulator is 300-nm-thick polycrystalline Y$_2$O$_3$ ($\varepsilon_r$ = 20). The SnO$_2$ channel is placed on the gap (~2 mm) between two Peltier devices, which are used to give temperature difference ($\Delta T$) between both edges of the channel. Two K-type thermocouples (T.C.) are located at both edges of the SnO$_2$ channel to measure the temperature difference between both edges of the SnO$_2$ channel. The thermo-electromotive force ($\Delta V$) and $\Delta T$ are measured simultaneously at fixed gate voltage ($V_g$).

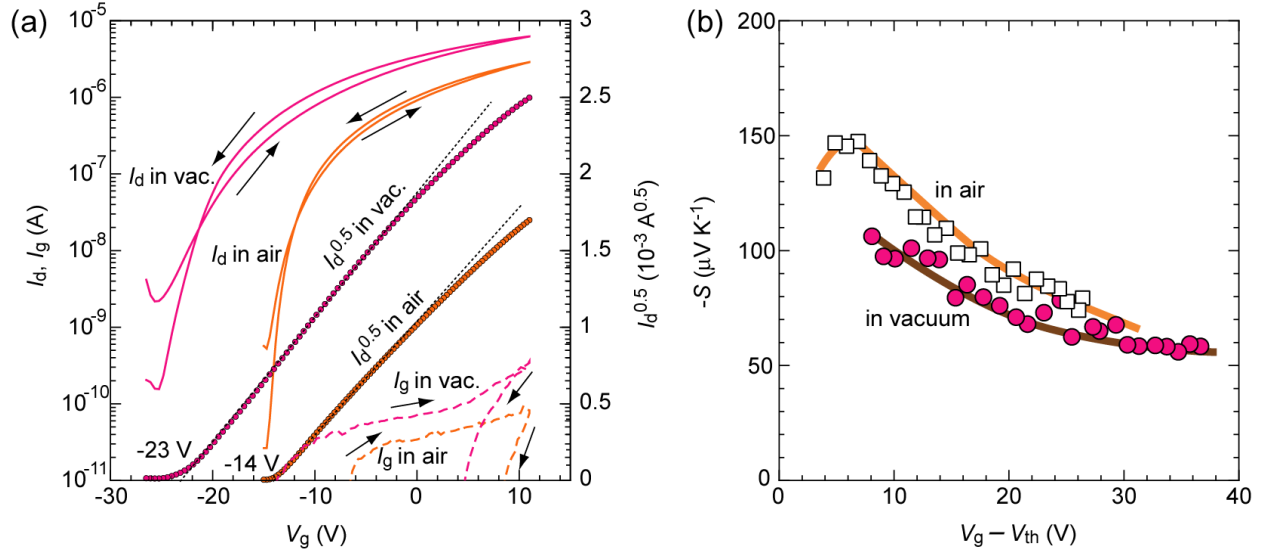

**FIG. 2** | Transistor characteristics the bottom-gate top-contact a-SnO$_2$ TTFT measured in air and in vacuum. (a) Transfer characteristic ($I_d$–$V_g$) curve at $V_d$ = +0.1 V. Corresponding $I_d^{0.5}$–$V_g$ and $I_g$–$V_g$ curves are also shown. The threshold voltage ($V_{th}$) is −14 V in air and −23 V in vacuum. The gate leakage current ($I_g$) is < 100 pA. (b) Electric field modulated thermopower ($S$) at various $V_g$–$V_{th}$ ranging from +4 V to +37 V. The −$S$ gradually decreases with $V_g$–$V_{th}$.

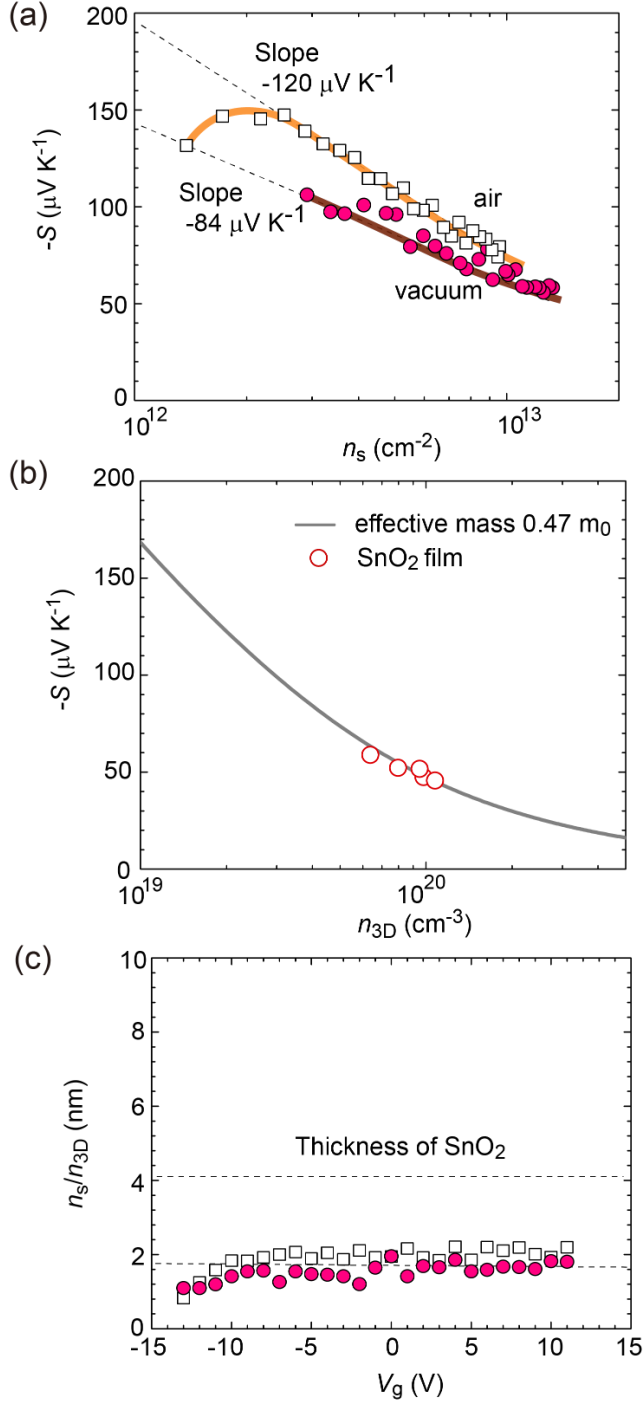

**FIG. 3** | Electric field thermopower modulation analyses of the bottom-gate top-contact a-SnO$_2$ TTFT. (a) Change in −$S$ as a function of the sheet carrier concentration ($n_s$). The slope of the −$S$−$n_s$ relationship is −120 μV K$^{-1}$ decade$^{-1}$ in air and −84 μV K$^{-1}$ decade$^{-1}$ in vacuum (dotted line). (b) three-dimensional carrier concentration ($n_{3D}$) dependence $S$ of the SnO$_2$ films. We calculated the carrier effective mass ($m^*$) of the SnO$_2$ film around 0.47 $m_0$. (c) The effective thickness ($t_{eff}$), which is defined as $n_s/n_{3D}$, as a function of $V_g$. The $t_{eff}$ is always ~1.7 ±0.4 nm, insensitive to the atmosphere.

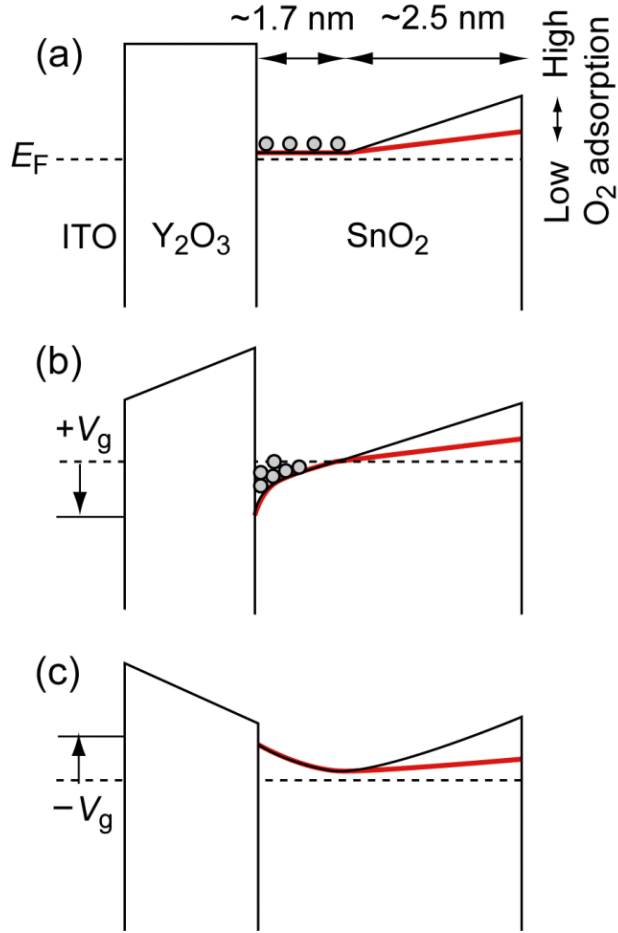

**FIG. 4** | Electric field modulation mechanism of the bottom-gate top-contact a-SnO$_2$ TTFT. (a) Without any gate voltage ($V_g$) application. Conduction band minimum (CBM) around the surface is lifted due to oxygen gas adsorption. Black: higher oxygen atmosphere (air). Red: lower oxygen atmosphere (vacuum). The 2.5-nm-thick surface region is the surface depletion layer. The conducting channel (~1 nm) remains at the Y$_2$O$_3$/SnO$_2$ interface. The sheet carrier concentration ($n_s$) is ~4 × 10$^{12}$ cm$^{-2}$. (b) Under positive $V_g$ application. The 1.7-nm-thick 2DEG layer is formed at the Y$_2$O$_3$/SnO$_2$ interface. The $n_s$ increases up to ~1 × 10$^{13}$ cm$^{-2}$. The surface region does not change. (c) Under negative $V_g$ application. The interface electrons are completely depleted, resulting in off states.

# Supplementary Material

# Electric field thermopower modulation analyses of the operation mechanism of transparent amorphous $SnO_2$ thin film transistor

Dou-dou Liang[1,2,a)], Yu-qiao Zhang[2], Hai Jun Cho[2], and Hiromichi Ohta[2,a)]

## AFFILIATIONS

[1]The Beijing Municipal Key Laboratory of New Energy Materials and Technologies, School of Materials Science and Engineering, University of Science and Technology Beijing, Beijing 100083, China

[2]Research Institute for Electronic Science, Hokkaido University, N20W10, Kita, Sapporo 001−0020, Japan

[a)]**Author to whom correspondence should be addressed:** liangdoudou1993@foxmail.com, hiromichi.ohta@es.hokudai.ac.jp

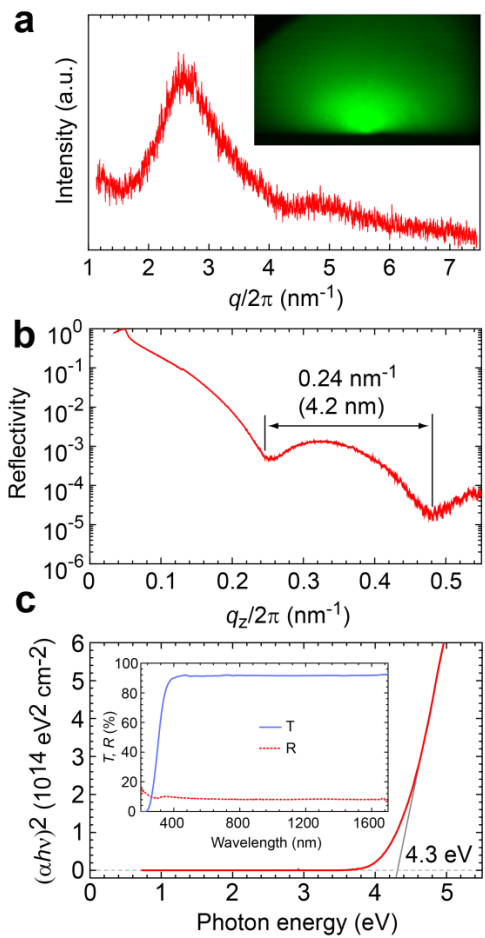

**FIG. S1** | Crystallographic analyses of the $SnO_2$ thin film. (a) Glancing angle (0.5°) incidence X-ray diffraction pattern. The inset shows the corresponding RHEED pattern. Only halo patterns are seen. (b) X-ray reflectivity of the $SnO_2$ thin film. The film thickness was evaluated as 4.2 nm. (c) Optical absorption spectrum (Tauc plot) of the $SnO_2$ thin film. The optical bandgap is 4.3 eV. Inset shows the transmission (T) and the reflection (R) spectra of the film.

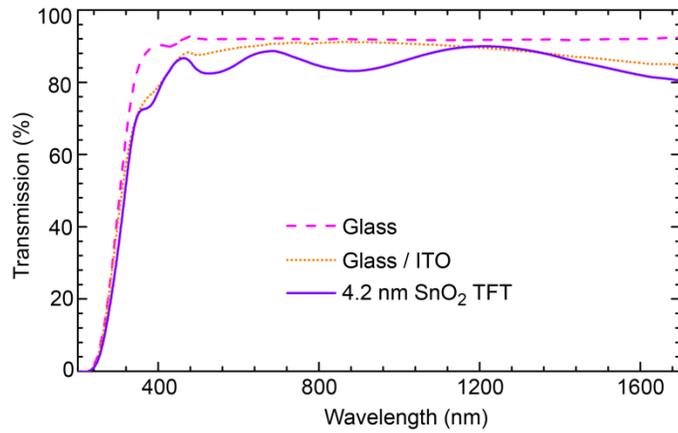

**FIG. S2** | Optical transmission of the bottom-gate top-contact a-SnO$_2$ TTFT. Optical transmission spectrum of the multiple layer composed of the 4.2-nm-thick a-SnO$_2$, 300-nm-thick polycrystalline Y$_2$O$_3$, and 100-nm-thick ITO on an Corning® EAGLE XG® (0.7-mm-thick) alkali-free glass substrate. The TTFT is fully transparent in the visible light region.

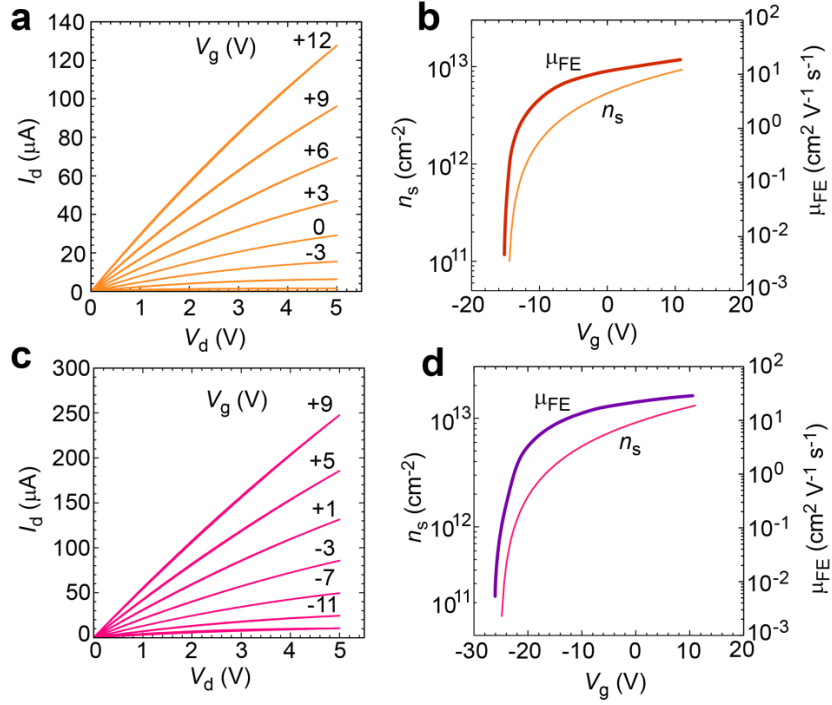

**FIG. S3** | Transistor characteristics of the bottom-gate top-contact a-SnO$_2$ TTFT. (a) Output characteristic ($I_d$–$V_d$) curves measured in air at various $V_g$ ranging from −9 V to +12 V. (b) Changes in the field effect mobility ($\mu_{FE}$) and the sheet carrier concentration ($n_s$) as function of $V_g$ measured in air. The $\mu_{FE}$ reaches ~20 cm$^2$ V$^{-1}$ s$^{-1}$. (c) $I_d$–$V_d$ curves measured in vacuum at various $V_g$ ranging from −15 V to +9 V. (d) $\mu_{FE}$ and $n_s$ as function of $V_g$ measured in vacuum. The $\mu_{FE}$ reaches ~30 cm$^2$ V$^{-1}$ s$^{-1}$.

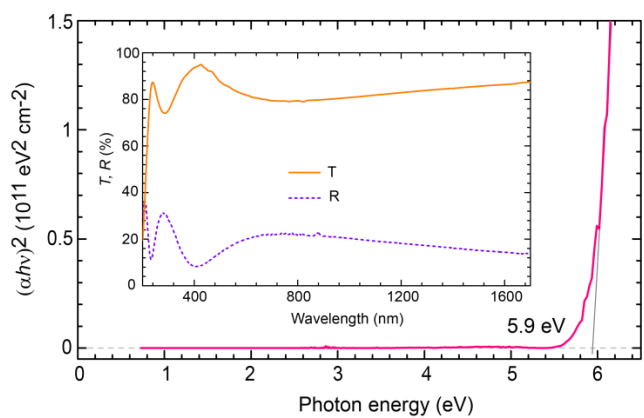

**FIG. S4** | Optical absorption spectrum of the polycrystalline $Y_2O_3$ thin film deposited on $SiO_2$ glass substrate. The optical bandgap is 5.9 eV. Inset shows transmission (T) and reflection (R) spectra. Note the thickness of the $Y_2O_3$ film was 104 nm, and the $Y_2O_3$ film was electrically insulator.

**Table S1**. Electrical properties of SnO$_2$ films measured at room temperature. We fabricated several SnO$_2$ films with different thickness ($t$) on glass substrates in the same manner of the main text. The resistivity ($\rho$), carrier concentration ($n$), and Hall mobility ($\mu_{Hall}$) of the films were measured using the conventional dc four-probe method with van der Pauw electrode geometry at room temperature. The thermopower ($S$) was acquired from the thermos-electromotive force ($\Delta V$) generated by a temperature difference ($\Delta T$) of ~15 K across the film using two Peltier devices. The temperatures at each end of the films were simultaneously measured with two thermocouples, and the $S$-values were calculated from the slope of the $\Delta T - \Delta V$ plots (correlation coefficient: $> 0.9999$).

| SnO$_2$ | $t$ (nm) | $\rho$ ($\Omega$ cm) | $n$ ($\times 10^{19}$ cm$^{-3}$) | $\mu_{Hall}$ (cm$^2$ V$^{-1}$ s$^{-1}$) | $-S$ ($\mu$V K$^{-1}$) |
|---|---|---|---|---|---|
| 1 | 4.50 | 5.35 | 7.98 | 14.61 | 52.12 |
| 2 | 5.88 | 4.43 | 9.80 | 14.38 | 47.37 |
| 3 | 6.51 | 4.45 | 10.75 | 13.05 | 45.56 |
| 4 | 9.27 | 7.79 | 9.49 | 8.44 | 53.08 |
| 5 | 11.2 | 11.90 | 6.38 | 8.23 | 58.80 |